\documentclass[reprint,amsmath,amssymb,aps,prl,floatfix]{revtex4-2}

\usepackage[T1]{fontenc}
\usepackage[utf8]{inputenc}
\usepackage{lmodern}

\usepackage{bm}
\usepackage{graphicx}
\usepackage{xcolor}
\usepackage{float}
\usepackage{enumitem}
\usepackage[export]{adjustbox}
\usepackage[
  protrusion=true,
  expansion=true,
  stretch=10,
  shrink=15,
  step=1,
  tracking=alltext, 
  letterspace=0
]{microtype}

\usepackage[unicode,colorlinks=true]{hyperref}

\DeclareMathOperator{\diag}{diag}
\DeclareMathOperator{\Tr}{Tr}
\newcommand{\TrA}{\operatorname{Tr}_{A}}
\newcommand{\TrB}{\operatorname{Tr}_{B}}

\begin{document}

\title{Quantum selective measurement as a quasilinear evolution
}

\author{Jakub Rembieliński}
\email{jaremb@uni.lodz.pl}
\author{Karol {\L}awniczak}
\affiliation{University of Lodz, Faculty of Physics and Applied Informatics,\\  
Pomorska 149/153, 90-236 Lodz, Poland}

\begin{abstract}
    We propose replacing the instantaneous state reduction in von Neumann selective measurement with continuous nonlinear evolution. Despite its nonlinearity, this evolution preserves the equivalence of quantum ensembles and hence obeys the no-signaling principle. Its final states coincide with those produced by the von Neumann projection. The defining features of rank-one projective measurement are retained: convergence to the eigenstate of the observable associated with the selected outcome, independence of this final state from the initial state, and consistent action on entangled states.
\end{abstract}

\maketitle

Selective measurement has traditionally been regarded as one of the foundational principles of quantum mechanics, as it provides an essential means of probing quantum reality~\cite{dEspagnat2003VeiledReality,
Schlosshauer2004RMP,
HanceHossenfelder2022JPhysCommun,
Albert2023GuessRiddle} and extracting information about the state of quantum systems. However, the status of quantum measurement remains conceptually ambiguous. A serious epistemic issue arises from the instantaneous nature of quantum state reduction (collapse). This is a long-standing conceptual problem in quantum mechanics, which, despite repeated attempts at its resolution, remains unresolved \cite{BassiEtAl2013RMP,BassiDoratoUlbricht2023Entropy,TomazMattosBarbatti2025PhilMag}.

Furthermore, selective measurement is based on a fundamentally different type of state transformation from the linear operations \cite{CarlessoEtAl2022NatPhys} that are typical in quantum mechanics. Specifically, it is evidently nonlinear in its deterministic component, namely in the von Neumann projection of the density matrix $\rho$,
\begin{equation}
\rho \mapsto \rho_{\mathrm{out}}=\frac{\Pi\rho\Pi}{\Tr{\left(\Pi\rho\right)}}\,,
\label{eq:von-neumann-update}
\end{equation}
where $\Pi$ is a projector from the spectral decomposition of the corresponding observable. 

Contrary to the common belief that nonlinearity destroys the equivalence of quantum ensembles, the von Neumann projection preserves ensemble equivalence. This means that selective measurement does not allow one to discriminate between ensembles corresponding to the same density operator. Recall that such discrimination is prohibited, as it would violate the no-signaling condition imposed by relativistic causality \cite{ZhanParis2014IJSI,Paris2012EPJST}.

A natural question, therefore, arises concerning the possible existence of other nonlinear operations that preserve quantum ensemble equivalence, notwithstanding the Gisin no-go theorem \cite{Gisin1989HPA,Gisin1990PhysLettA,SimonBuzekGisin2001PRL}, according to which nonlinear quantum dynamics would lead to superluminal signaling. However, a number of limitations on the applicability of this powerful theorem have been identified (see, e.g., \cite{Czachor1998PRA,CzachorDoebner2002PhysLettA,Kent2005PRA,HelouChen2017JPCS}). An exhaustive analysis of this issue from the perspective of recent developments is given in \cite{BielinskaEcksteinHorodecki2025NJP}.

Moreover, recent papers \cite{RembielinskiCaban2020PRR,RembielinskiCaban2021Quantum,RembielinskiCiborowski2023AOP} have introduced the notion of quasilinear operations and evolutions, which are nonlinear yet simultaneously preserve both the equivalence of quantum ensembles and the convex structure of the set of density operators. Consequently, superluminal communication is forbidden in their case. The class of such operations includes the von Neumann projections, \eqref{eq:von-neumann-update}. Therefore, quasilinear operations and evolutions, although nonlinear, can be regarded as legitimate quantum-mechanical operations.

In this Letter, we show that replacing the von Neumann projection with a quasilinear evolution does not change the final results of selective measurement. This Letter presents a minimal self-contained construction and its basic two-level implementation; detailed derivations, further analysis of the dynamics, some extensions, and the application to the specific physical experiment are given in the companion article~\cite{AssociatedLong}. As we demonstrate using the simplest two-level quantum system, quasilinear evolution, when replacing the von Neumann projection, reproduces all its essential properties, including the correct action on entangled states. This leads to the following conjecture: if measurement is accepted as a quasilinear process over time, then the projection postulate simply expresses the final state of this evolution.

Let us begin with a brief and simplified description of quasilinear operations introduced in \cite{RembielinskiCaban2020PRR,RembielinskiCaban2021Quantum,RembielinskiCiborowski2023AOP}. A precise and exhaustive discussion of convex quasilinear maps and quasilinear evolution is given in \cite{AssociatedLong}. Below, the underlying Hilbert space is denoted by $\mathcal{H}$ while the convex set of density matrices by $S(\mathcal{H})$.

\vspace{1ex}
\noindent Quasilinear map --- A map $\Phi:S(\mathcal{H})\to S(\mathcal{H})$ on the manifold of density matrices $S(\mathcal{H})$ is convex quasilinear if for all ensembles
$\rho=\varepsilon\,\rho_{a}+(1-\varepsilon)\,\rho_{b}\in S(\mathcal{H})$ and $\varepsilon\in[0,1]$, there exists $\varepsilon'\in[0,1]$ such that
\begin{equation}
\rho'=\Phi(\rho)=\Phi\left(\varepsilon\,\rho_a+(1-\varepsilon)\,\rho_b\right)
=\varepsilon'\,\Phi(\rho_a)+(1-\varepsilon')\,\Phi(\rho_b)\,.
\label{eq:quasilinear-map}
\end{equation}

It is evident that condition \eqref{eq:quasilinear-map} is satisfied for all linear maps, with $\varepsilon'=\varepsilon$ in that case. Thus, linear maps belong to the class of quasilinear operations. It is easy to see that the von Neumann map \eqref{eq:von-neumann-update} also satisfies the condition \eqref{eq:quasilinear-map} with
$\varepsilon'=\varepsilon\Tr{\Pi\rho_a}/\Tr{\Pi\rho}$.
Furthermore, a quasilinear map preserves equivalence of ensembles. 
If $\rho$ is represented by two different ensembles,
\begin{subequations}\label{eq:ensemble-equivalence}
\begin{align}
\rho&=\varepsilon\,\rho_a+(1-\varepsilon)\,\rho_b
=\widetilde{\varepsilon}\,\widetilde{\rho}_a+(1-\widetilde{\varepsilon})\,\widetilde{\rho}_b\,,
\label{eq:ensemble-equivalence-a}
\intertext{they are equivalent. After the action of a quasilinear operation $\Phi$, we obtain as a result that}
\rho'&=\Phi(\rho)=\varepsilon'\,\rho_a'+(1-\varepsilon')\,\rho_b'
=\widetilde{\varepsilon}'\,\widetilde{\rho}_a'+(1-\widetilde{\varepsilon}')\,\widetilde{\rho}_b'\,,
\label{eq:ensemble-equivalence-b}
\end{align}
\end{subequations}
and $0\le \varepsilon'\le 1$ and $0\le \widetilde{\varepsilon}'\le 1$. Therefore, the new ensembles are equivalent as well.

The notion of convex quasi-linearity can be extended to deterministic time evolutions realized as a one-parameter dynamical semigroup $\Phi_t$ on the convex set of density operators $S(\mathcal{H})$. It was introduced in \cite{RembielinskiCaban2021Quantum}, where the quasilinear extension of the Gorini--Kossakowski--Sudarshan--Lindblad (GKSL) equation \cite{GoriniKossakowskiSudarshan1976JMP,Lindblad1976CMP} was derived.

\noindent Quasilinear evolution --- An evolution $\Phi_t$ is convex quasilinear if for all density operators from $S(\mathcal{H})$, $\Phi_t$ satisfies the following condition:
\begin{equation}
\begin{split}
\rho(t)
&=\Phi_t(\rho_0) \\
&=\Phi_t\left(\varepsilon_0\rho_{a0}+(1-\varepsilon_0)\rho_{b0}\right) \\
&=\varepsilon(t)\Phi_t(\rho_{a0})+
\left(1-\varepsilon(t)\right)\Phi_t(\rho_{b0}) \\
&=\varepsilon(t)\rho_a(t)+\left(1-\varepsilon(t)\right)\rho_b(t)\,,
\end{split}
\label{eq:quasilinear-evolution-mixture}
\end{equation}
provided that for each $t$, $0\le\varepsilon(t)\le 1$ holds. Here $\rho_0=\varepsilon_0\,\rho_{a0}+(1-\varepsilon_0)\,\rho_{b0}$ is the initial density matrix.
Thus, quasilinear evolution $\Phi_t$ transforms an initial convex combination $\varepsilon_0\,\rho_{a0}+(1-\varepsilon_0)\,\rho_{b0}$ given at the time $t_0=0$ into a convex combination $\varepsilon(t)\,\rho_a(t)+(1-\varepsilon(t))\,\rho_b(t)$ at the time $t$. Parenthetically, linear evolutions form a subfamily in this class. A discussion and examples of quasilinear evolutions are given in \cite{RembielinskiCaban2021Quantum}.
In contrast to the general nonlinear evolution case, the convex quasilinear map preserves the convex structure of the set of quantum ensembles and the equivalence of ensembles \cite{AssociatedLong}. 

The simplest quasilinear GKSL equation, being a nonlinear generalization of the von Neumann equation, is of the form
\begin{equation}
\hbar\,\frac{d}{dt}\rho
=-i[H,\rho]+\{G,\rho\}-2\rho\,\Tr{\left(G\rho\right)}\,,
\label{eq:quasilinear-gksl}
\end{equation}
where $H$ is the Hamiltonian, while the Hermitian operator $G$ constitutes the nonlinear part of \eqref{eq:quasilinear-gksl} and can be time dependent \cite{RembielinskiCaban2021Quantum}. The trace condition $\Tr{\rho}=1$ remains satisfied throughout the evolution \eqref{eq:quasilinear-gksl}, ensuring probability conservation. Moreover, the purity condition  $\rho^2=\rho$ is also preserved by this evolution, which means that pure states evolve into pure states; see \cite{AssociatedLong}. 

In general, the energy of the system described by the Hamiltonian $H$ need not be conserved. Thus, the system is not isolated. The operator $G$ represents the effective influence of the environment on the system's state.

\vspace{1ex}
\paragraph{Quasilinear selective measurement ---}
In this Letter, we consider a situation in which the deterministic part of the selective measurement (projection) is replaced by a quasilinear evolution of a quantum system, as given in \eqref{eq:quasilinear-gksl}. The generator $H$ will be reinterpreted as a quantum observable $\Omega$, which we take to be a time-independent operator. On the other hand, the effective influence of the measuring apparatus on the state of the system is accounted for by the generator $G$, which is, in general, time-dependent. With these generators, we adopt the quasilinear evolution in the form
\begin{equation}
\hbar\,\frac{d}{dt}\rho
=-i[\Omega,\rho]+\{G(t),\rho\}-2\rho\,\Tr{G(t)\rho}\,,
\label{eq:measurement-quasilinear-eq}
\end{equation}
with the initial condition $\rho(0)=\rho_0$. Naturally, $\Omega$ and $G$ should be given in energy units.

It is not difficult to show that the solution $\rho(t)$ of \eqref{eq:measurement-quasilinear-eq} can be obtained in the Kraus-like time-dependent form. Namely, as is shown in \cite{AssociatedLong},
\begin{equation}
\rho(t)=\frac{K(t)\rho_0K^{\dagger}(t)}{\Tr{\left(K(t)\rho_0K^{\dagger}(t)\right)}}\,,
\label{eq:kraus-solution}
\end{equation}
where $K(t)$ is related to the quasilinear von Neumann equation \eqref{eq:measurement-quasilinear-eq} via the linear differential equation
\begin{equation}
i\hbar\frac{d}{dt}K(t)-\left(\Omega+iG(t)\right)K(t)=0,
\qquad K(0)=I\,.
\label{eq:K-linear-eq}
\end{equation}
Equation~\eqref{eq:K-linear-eq} is state independent.
Using $\rho(t)=\Phi_t(\rho_0)$ defined by \eqref{eq:quasilinear-evolution-mixture} in \eqref{eq:kraus-solution} together with the relation \eqref{eq:K-linear-eq} leads to the conclusion that the evolution \eqref{eq:kraus-solution} (so \eqref{eq:measurement-quasilinear-eq} too) is quasilinear with the coefficient $\varepsilon(t)$ of the form
\begin{equation}
\varepsilon(t)=\varepsilon_0\,
\frac{\Tr{\left(K(t)\rho_{a0}K^{\dagger}(t)\right)}}{\Tr{\left(K(t)\rho_0K^{\dagger}(t)\right)}},
\qquad \varepsilon(0)=\varepsilon_0\,.
\label{eq:epsilon-evolution}
\end{equation}

\vspace{1ex}
\paragraph{Selective measurement as quasilinear state evolution of a two-level quantum system ---}
Now, we specialize the construction to a general two-level system and compare the quasilinear selective measurement defined by \eqref{eq:measurement-quasilinear-eq}, or equivalently by \eqref{eq:kraus-solution} and \eqref{eq:K-linear-eq}, with the von Neumann projective measurement. To this end, we must specify an observable $\Omega$ and a driving generator $G(t)$ that accounts for the influence of the apparatus on the system. The observable $\Omega$ is represented by
\begin{equation}
\Omega(\bm{\omega})=\frac{1}{2}\,\bm{\omega}\cdot\bm{\sigma}\,,
\label{eq:observable-omega}
\end{equation}
where $\bm{\omega}$ is a time-independent real vector and $\bm{\sigma}$ denotes the triplet of Pauli matrices. The spectral decomposition of $\Omega(\bm{\omega})$ is of the form
\begin{equation}
\Omega(\bm{\omega})=\frac{\omega}{2}\left(\Pi_{+}(\bm{\omega})-\Pi_{-}(\bm{\omega})\right),
\quad
\Pi_{\lambda}(\bm{\omega})=\frac{1}{2}\left(I+\lambda\,\hat{\bm{\omega}}\cdot\bm{\sigma}\right),
\label{eq:omega-spectral-decomp}
\end{equation}
where $\omega=|\bm{\omega}|$, $\hat{\bm{\omega}}=\bm{\omega}/\omega$, $\lambda=\pm 1$, and $\lambda\omega/2$ are the eigenvalues of $\Omega$. The driving generator is consequently expressed in the form
\begin{equation}
G_{\lambda}(t)=\frac{\lambda}{2}\,\bm{g}(t)\cdot\bm{\sigma}\,,
\label{eq:driving-generator}
\end{equation}
The generator $G_\lambda(t)$ is understood as an effective generator on the conditional branch with $\lambda=\pm 1$ denoting the outcome selected in a selective measurement. Its probability is given by the Born rule. Once the branch $\lambda$ has been selected, the subsequent evolution is deterministic.

In the corresponding Bloch representation, the density operator $\rho$ takes the form
\begin{equation}
\rho=\frac{1}{2}\left(I+\bm{n}\cdot\bm{\sigma}\right),
\qquad |\bm{n}|^{2}\le 1\,.
\label{eq:bloch-density}
\end{equation}
In particular, projectors $\Pi_{\lambda}(\bm{\omega})$ correspond to the Bloch vectors $\bm{\pi}_{\lambda}=\lambda\hat{\bm{\omega}}$.

In terms of $\bm{\omega}$, $\bm{g}(t)$ and the Bloch vector $\bm{n}(t)$, the quasilinear evolution equation \eqref{eq:measurement-quasilinear-eq} takes the following form:
\begin{equation}
\hbar\,\frac{d}{dt}\bm{n}(t)
=\bm{\omega}\times\bm{n}(t)
+\lambda\,\bm{g}(t)
-\lambda\left(\bm{g}(t)\cdot\bm{n}(t)\right)\bm{n}(t).
\label{eq:bloch-quasilinear-eq}
\end{equation}
The scalar multiplication of \eqref{eq:bloch-quasilinear-eq} with $\bm{n}(t)$ leads to
\begin{equation}
\frac{\hbar}{2}\frac{d}{dt}|\bm{n}(t)|^{2}
=\lambda\bm{g}(t)\cdot\bm{n}(t)\left(1-|\bm{n}(t)|^{2}\right),
\label{eq:bloch-norm-eq}
\end{equation}
so the purity condition $\rho^{2}=\rho$, i.e.\ $|\bm{n}(t)|^{2}=1$, is preserved under the quasilinear evolution. Equation \eqref{eq:bloch-quasilinear-eq} forms a Riccati system belonging to the class of structurally unstable differential equations \cite{ArnoldEtAl1994DynamicalSystemsV}. This instability consists of a radical, qualitative change in the nature of the evolution within a narrow region of the equations' parameter space. In the present model, the relevant critical configurations are associated with $\bm{\omega}\perp\bm g$ and with trajectory points at which $\omega^2=g(t)^2$.

\begin{figure}[H]
    \centering
    \begin{tabular}{c c}
        \includegraphics[width=0.24\textwidth]{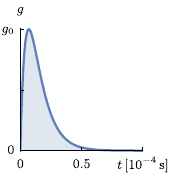} & \includegraphics[width=0.24\textwidth]{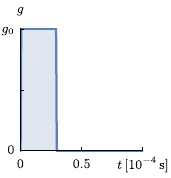} \\
        Inverted Morse (IM) & Constant over an interval \\
        \end{tabular}
\caption{Forms of potential considered.}
\label{fig:driving-potentials}
\end{figure}

In the following, we employ the polar parametrization of the vectors 
$\bm{\omega} = \omega\left(\sin\alpha\cos\beta,\sin\alpha\sin\beta,\cos\alpha\right)$, $\bm{g}(t) = g(t)\left(\sin\theta\cos\varphi,\sin\theta\sin\varphi,\cos\theta\right)$.
For simplicity, we assume that the time dependence of $\bm{g}(t)$ resides solely in its magnitude $g(t)$, while its direction remains time-independent. As follows from our requirements discussed below, the vector $\bm{g}(t)$, which governs the measurement process, is zero before the measurement and decays to zero afterward. Consequently, its magnitude $g(t)$ must attain a maximum during the measurement interval. In this Letter, we present the results based on the two representative profiles $g(t)$: an inverted Morse (IM) potential
\begin{subequations}\label{eq:driving-profile}
\begin{align}
g(t)&=g_{0}\left(1-\left(1-2e^{-\kappa t}\right)^{2}\right),
\label{eq:driving-morse}\\
\intertext{and a potential constant over an interval}
g(t)&=g_{0}\,\mathcal{S}(t),
\label{eq:driving-window}
\end{align}
\end{subequations}
where $\mathcal{S}(t)$ is a smooth window function supported on a finite interval (see Fig.~\ref{fig:driving-potentials}).

The specific form of the time-dependent vector $\bm{g}(t)$ is determined by the configuration of the measuring device. For instance, in the Stern--Gerlach experiment, the orientation angles $\theta$ and $\varphi$, associated with the magnetic field configuration, remain constant. However, in general, the orientation of $\bm{g}(t)$ may also vary with time.

We demonstrate that the quasilinear evolution exhibits several properties characteristic to quantum selective measurements implemented with rank-one projectors, namely:
\begin{itemize}[noitemsep,leftmargin=*]
\item It involves a stochastic (random) choice of the outcome, followed by a deterministic update of the quantum state.
\item The final state of the evolution is independent of the initial state.
\item As a result of the measurement, the system reaches a pure state that is an eigenstate of the observable $\Omega$.
\item The pre-measurement state affects only the probabilities of possible outcomes, as determined by the spectral decomposition of the observable and by the Born rule.
\item Selective measurement is well defined on the tensor-product states.
\end{itemize}
The last property is important given that nonlinear operations are often not well defined for entangled states \cite{Czachor1998PRA,BielinskaEcksteinHorodecki2025NJP,Mielnik2001PhysLettA}. We impose two additional, rather natural requirements:
\begin{itemize}[noitemsep,leftmargin=*]
\item The action of the generator $G$ should vanish (at least asymptotically) at the beginning and at the end of the measurement interval, which is assumed to be finite in spacetime. This requirement restricts the class of admissible functions $g(t)$.
\item The final state of the evolution should be insensitive to admissible variations in the form of the generator $G$, which accounts for the influence of the apparatus on the system.
\end{itemize}
The randomness inherent in both quasilinear and von Neumann measurements arises from the same stochastic postselection process that incorporates the Born rule.

Below, we present the results of numerical calculations of the evolution of Bloch vectors governed by \eqref{eq:measurement-quasilinear-eq} in the form \eqref{eq:bloch-quasilinear-eq} or, equivalently, by its global form \eqref{eq:kraus-solution}, \eqref{eq:K-linear-eq}.

\begin{widetext}

\begin{figure}[H]
    \centering
    \begin{tabular}{c c c}
        & \makebox[0.4375\textwidth][c]{$\lambda=+1$}
            \makebox[0.4375\textwidth][c]{$\lambda=-1$}
        & \\
        & \includegraphics[valign=c,width=0.875\textwidth]{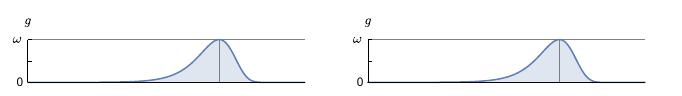} & \\
        & \includegraphics[valign=c,width=0.875\textwidth]{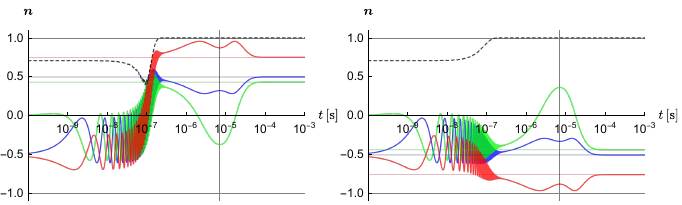} & \\
        & \includegraphics[valign=c,width=0.750\textwidth]{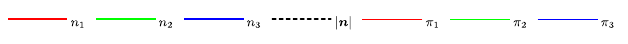} & \\
        & \includegraphics[valign=c,width=0.875\textwidth]{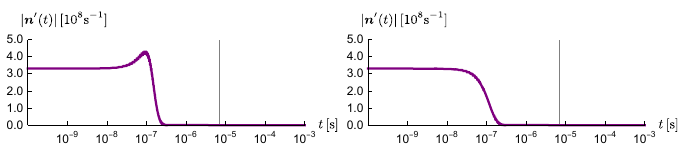} & \\
        \end{tabular}
    \caption{Evolution of the state. The vector defining the observable, directed at $\alpha = \pi/3$, $\beta = \pi/6$ has the following components $\bm{\omega}=\omega\bigl[\frac{3}{4},\frac{\sqrt{3}}{4},\frac{1}{2}\bigr]$, with $\omega=10^{9}\hbar/s$. The vector defining the nonlinear part of the evolution generator is directed at $\theta=\frac{\pi}{6}$, $\varphi=-\frac{\pi}{3}$, so $\bm{g}=g(t)\bigl[\frac{1}{4},-\frac{\sqrt{3}}{4},\frac{\sqrt{3}}{2}\bigr]$. Its magnitude $g(t)$ is chosen according to the inverted Morse potential, \eqref{eq:driving-morse}, with parameter $\kappa=10^5 /s$ and peak value $g_{0}=10^{9}\hbar/s$. The form of the potential is depicted above the main plot. The angle $\Theta\approx1.12$ between $\bm{\omega}$ and $\bm{g}$ is significantly smaller than $\pi/2$. The initial state $\bm{n}_0$ is chosen to be a state of intermediate purity $\bm{n}_{0}=\frac{1}{\sqrt{2}}\bigl[-\frac{1}{\sqrt{2}},0,-\frac{1}{\sqrt{2}}\bigr]$. The two columns show the dynamics conditioned on the two possible measurement outcomes: $\lambda=+1$ and $\lambda=-1$. In each plot, the horizontal axis indicates the time $t$ on a logarithmic scale. The red, green, and blue lines denote the components of the Bloch vector $\bm{n}(t)$: $n_x$, $n_y$, and $n_z$, respectively; the thin straight lines in the same colors denote the respective components of the vector $\bm{\pi}_\lambda$ representing the state resulting from the von Neumann projection. The black dashed line is the norm $|\bm{n}(t)|$ of the vector $\bm{n}(t)$. Below, with a purple line, the rate of change $|\bm{n}'(t)|$ of the vector $\bm{n}(t)$ over time is presented. The position of the potential's maximum is indicated by the vertical lines.}
    \label{fig:state-evolution}
\end{figure}
\end{widetext}

\vspace{1ex}
\paragraph{Comparison of the von Neumann and quasilinear selective measurements. Numerical results ---} 
We now use \eqref{eq:bloch-quasilinear-eq} to demonstrate the properties of quasilinear measurement and compare it with the von Neumann projection. Figure \ref{fig:state-evolution} illustrates an example of such an evolution under an IM potential on a time scale comparable to that of the Stern--Gerlach experiment, discussed in \cite{AssociatedLong}. The parameters were selected to provide a representative example. A broader parameter-space analysis, including the critical region, is given in the companion article~\cite{AssociatedLong}. Cases of evolution conditioned on both possible measurement outcomes $\lambda=\pm 1$ are presented side by side.

The analysis of the state's evolution, presented in Fig.~\ref{fig:state-evolution}, as well as in other cases studied, allows us to distinguish two stages of this process.

\noindent Early phase of evolution --- Initially, the Bloch vector components undergo oscillations. Then, they enter a damping stage, which lasts until the oscillations disappear. Simultaneously, the initially mixed state turns into a pure state. After this, the rate of change of the Bloch vector drops to almost zero.
\newpage

\noindent Late phase of evolution --- Subsequently, a transient disturbance of the state is observed. Ultimately, the Bloch vector attains its final state, which is the eigenstate of the observable $\Omega$, represented by the vectors $\bm{\pi}_\lambda=\lambda\hat{\bm{\omega}}$, with $\lambda=\pm 1$ labeling the corresponding eigenvalue.

The initial phase of the evolution can be interpreted as dominated by the evolution generator $\Omega(\bm{\omega})$, while the influence of the generator $G(\bm{g})$ remains relatively weak (see the form of the potential). This leads to precession of the Bloch vector $\bm{n}(t)$. As the potential increases, the influence of the generator $G$ becomes stronger, resulting in the damping of the Bloch vector's precession. 

\begin{widetext}

\begin{figure}[H]
    \centering
    \begin{tabular}{c c c}
        & \makebox[0.4375\textwidth][c]{$\lambda=+1$}
            \makebox[0.4375\textwidth][c]{$\lambda=-1$}
        & \\
         \rotatebox[origin=c]{90}{$\bm{n}_{0}=\bigl[\frac{3}{4},-\frac{\sqrt{3}}{4},-\frac{1}{2}\bigr]$} & \includegraphics[valign=c,width=0.875\textwidth]{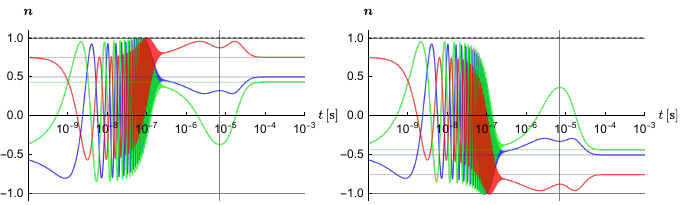} & \\
         \rotatebox[origin=c]{90}{$\bm{n}_0=\bm{0}$} & \includegraphics[valign=c,width=0.875\textwidth]{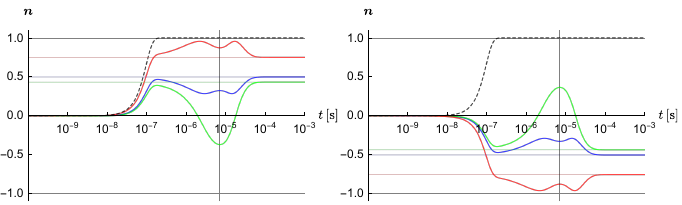} & \\
         & \includegraphics[valign=c,width=0.750\textwidth]{legend.pdf} & \\
        \end{tabular}
    \caption{Evolution of the state from two further (following Fig.~\ref{fig:state-evolution}) initial states $\bm{n}_0$: one pure and one fully depolarized. The vectors $\bm{\omega}$ and $\bm{g}$, and hence $\Theta$, are kept the same as in Fig.~\ref{fig:state-evolution}.}
    \label{fig:initial-state-independence}
\end{figure}
\end{widetext}

\begin{widetext}

\begin{figure}[H]
    \centering
    \begin{tabular}{c c c}
        & \makebox[0.4375\textwidth][c]{$\lambda=+1$}
            \makebox[0.4375\textwidth][c]{$\lambda=-1$}
        & \\
         & \includegraphics[valign=c,width=0.875\textwidth]{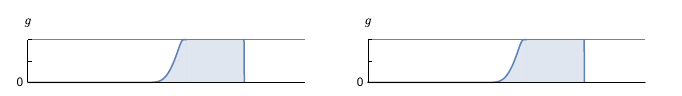} & \\
         & \includegraphics[valign=c,width=0.875\textwidth]{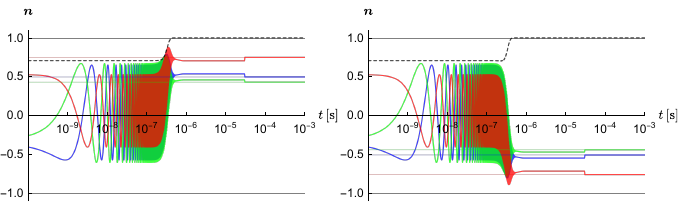} & \\
         & \includegraphics[valign=c,width=0.750\textwidth]{legend.pdf} & \\
        \end{tabular}
    \caption{State dynamics for a substantially different angle $\Theta$ between vectors $\bm{g}$ and $\bm{\omega}$ and a different function $g(t)$ with a different scale factor $g_0$. Vector $\bm{\omega}$ is kept the same as in Fig.~\ref{fig:state-evolution}, however, the orientation $\hat{\bm{g}}=\bigl[\frac{1}{2},\frac{\sqrt{3}}{2},0\bigr]$ is different. The angle $\Theta\approx 0.723$ is still significantly smaller than $\pi/2$. The form of the function $g(t)$ is completely different -- constant over an interval, \eqref{eq:driving-window}, shown above the main plot. Its maximal value $g_{0}=10^8\hbar/s$ is also very different from that in the previous cases. The initial state $\bm{n}_{0}=\frac{1}{\sqrt{2}}\bigl[\frac{3}{4},-\frac{\sqrt{3}}{4},-\frac{1}{2}\bigr]$.}
    \label{fig:robustness-form}
\end{figure}
\end{widetext}

As shown in Fig.~\ref{fig:initial-state-independence}, this oscillatory phase is not present in the case of the maximally mixed initial state. However, rapid conversion of a mixed state into a pure state is unavoidable and constitutes a key element of the process of transforming an arbitrary state into an eigenstate of the observable. As illustrated in Fig.~\ref{fig:robustness-form}, this disturbance can be weakened by lowering the peak magnitude $g_0$ of the vector $\bm{g}(t)$ or eliminated by setting the vector $\bm{g}(t)$ parallel to $\bm{\omega}$, that is, by suitable adjustment of the experimental apparatus.

\vspace{1ex}
\paragraph{Independence from the initial state ---}
The final state should be independent of the initial state. The following figure shows examples of evolution under the IM potential starting from completely different initial states.

In the examples shown in Figs.~\ref{fig:state-evolution} and~\ref{fig:initial-state-independence}, the asymptotic state is identical. In all cases considered, the final state of the evolution is independent of the initial state, thereby satisfying the requirement imposed on the measurement dynamics.

\paragraph{Independence from variations of the potential $\bm{g}$ ---}
The final state should be resistant to a wide class of modifications of the potential. Firstly, $\bm{g}(t)$ can have a different scale (quantified with $g_0$). Secondly, the orientation of the vector $\bm{g}$ may be changed. What is important in $\bm{g}$ orientation is its angle of deviation from the $\bm{\omega}$ vector: $\Theta$. Thirdly, the form of the $g(t)$ function may be chosen in various ways. Of course, there is nothing special about the IM potential, and it may be replaced with a completely different one, provided that it decays to zero after the time of the measurement, at least asymptotically. In the following figure, we present the state dynamics in the case of a completely different scale $g_0$, orientation $\hat{\bm{g}}$, and form of the function $g(t)$.

The final state is independent of the scale factor $g_0$ over a wide range of its values and of the orientation of $\bm{g}$ as far as the angle $\Theta$ between $\bm{g}$ and $\bm{\omega}$ does not approach the critical value of $\pi/2$, at which the structural instability of the equations manifests itself. The behavior of the state when $\Theta \approx \pi/2$ and $\Theta>\pi/2$ is the matter of the article \cite{AssociatedLong}. 
The form of $g(t)$ does not influence the final state, provided that the integrated strength of the driving term $\int{g(t)\,dt}$ is sufficiently large and $\bm g$ remains away from the critical region. These three features of the final state's robustness with respect to variations in the potential $\bm{g}$ satisfy the conditions imposed on the dynamics. 

Indeed, the asymptotic final state is independent of the specific form of the driving potential. 
Moreover, as we demonstrate in \cite{AssociatedLong}, when the driving term satisfies the conditions mentioned above, the global evolution \eqref{eq:kraus-solution} yields the projection formula.

\paragraph{Local quasilinear measurement on entangled states ---}
Because the evolution equation \eqref{eq:measurement-quasilinear-eq} is integrated to the global Kraus-like form \eqref{eq:kraus-solution}, local quasilinear measurements are well defined on entangled states. 

To show this, let us consider such a local action in the channel $A$ on the general state in the tensor product space $\mathcal{H}_A\otimes\mathcal{H}_B$ of two 2-dimensional quantum systems $A$ and $B$, namely on
\begin{equation}
\rho_{0}=\frac{1}{4}\left(I\otimes I+\bm{n}_{\mathrm{A}0}\cdot\bm{\sigma}\otimes I
+I\otimes \bm{n}_{\mathrm{B}0}\cdot\bm{\sigma}
+T_{ij}\,\sigma_{i}\otimes\sigma_{j}\right).
\label{eq:twoqubit-state}
\end{equation}

Using \eqref{eq:kraus-solution} and \eqref{eq:K-linear-eq}, we obtain that

\begin{equation}
\begin{split}
\rho_{\lambda}(t)
&=\frac{\bigl(K_{\lambda}(t)\otimes I\bigr)\rho_{0}\bigl(K_{\lambda}^{\dagger}(t)\otimes I\bigr)}
{\Tr{\bigl(K_{\lambda}(t)\otimes I\bigr)\rho_{0}\bigl(K_{\lambda}^{\dagger}(t)\otimes I\bigr)}}=\\
&=\frac{1}{4}\Biggl(
\frac{2\,K_{\lambda}(t)\rho_{\mathrm{A}0}K_{\lambda}^{\dagger}(t)}{\Tr{\bigl(K_{\lambda}(t)\rho_{\mathrm{A}0}K_{\lambda}^{\dagger}(t)\bigr)}}\otimes I+\\
&+\frac{K_{\lambda}(t)K_{\lambda}^{\dagger}(t)}{\Tr{\bigl(K_{\lambda}(t)\rho_{\mathrm{A}0}K_{\lambda}^{\dagger}(t)\bigr)}}\otimes\left(\bm{n}_{\mathrm{B}0}\cdot\bm{\sigma}\right)+\\
&+T_{ij}\,
\frac{K_{\lambda}(t)\sigma_{i}K_{\lambda}^{\dagger}(t)}{\Tr{\bigl(K_{\lambda}(t)\rho_{\mathrm{A}0}K_{\lambda}^{\dagger}(t)\bigr)}}\otimes\sigma_{j}
\Biggr).
\end{split}
\label{eq:postmeasurement-twoqubit}
\end{equation}    
The subscript $\lambda$ corresponds to two possible trajectories related to the measurement outcome in the channel $A$.

The local densities evolve according to
\begin{subequations}\label{eq:evolved-reduced-states}
    \begin{align}
&\rho_{\mathrm{A}\lambda}(t)
=\TrB{\rho_{\lambda}(t)}
=\frac{K_{\lambda}(t)\rho_{\mathrm{A}0}K_{\lambda}^{\dagger}(t)}{\Tr{\left(K_{\lambda}(t)\rho_{\mathrm{A}0}K_{\lambda}^{\dagger}(t)\right)}},
\label{eq:evolved-A}\\
\begin{split}
&\rho_{\mathrm{B}\lambda}(t)
=\TrA{\rho_{\lambda}(t)}=\frac{1}{2}\Biggl(
I+\\
&+\frac{\left(
n_{\mathrm{B}0j}\,\Tr{\left(K_{\lambda}(t)K_{\lambda}^{\dagger}(t)\right)}+T_{ij}\,\Tr{\left(K_{\lambda}(t)\sigma_{i}K_{\lambda}^{\dagger}(t)\right)}
\right)\sigma_{j}}{2\,\Tr{\left(K_{\lambda}(t)\rho_{\mathrm{A}0}K_{\lambda}^{\dagger}(t)\right)}}\Biggr).
\end{split}
\label{eq:evolved-B}
\end{align}
\end{subequations}

The standard projective measurement, by means of the von Neumann rule, leads to analogous formulas with $K_{\lambda}(t)$ replaced by the projectors $\Pi_{\lambda}$. 
Below, we give an explicit example of a quasilinear measurement with an entangled state.

\begin{widetext}

\begin{figure}[H]
    \centering
    \begin{tabular}{l c c}
        & \makebox[0.4375\textwidth][c]{$\lambda=+1$}
            \makebox[0.4375\textwidth][c]{$\lambda=-1$}
        & \\
        A & \includegraphics[valign=c,width=0.875\textwidth]{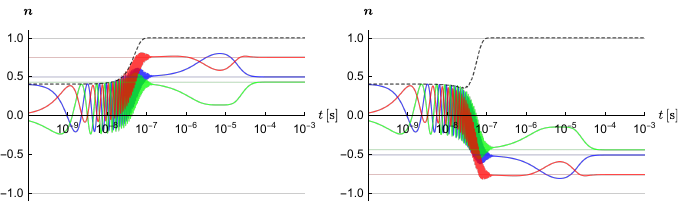} & \\
        B & \includegraphics[valign=c,width=0.875\textwidth]{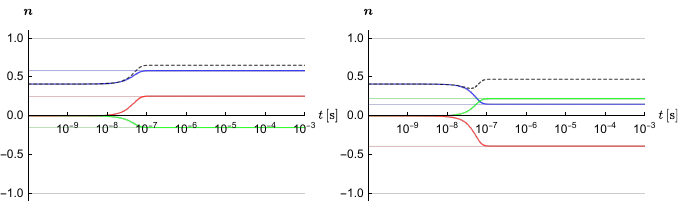} & \\
        & \includegraphics[valign=c,width=0.750\textwidth]{legend.pdf} & \\
        \end{tabular}
    \caption{The dynamics of subsystems $A$ and $B$ of the entangled system of two qubits subject to quasilinear measurement in the channel $A$. Vectors $\bm{\omega}$ and $\bm{g}$ (hence $\Theta$) are kept the same as in Fig.~\ref{fig:state-evolution}. The initial state is chosen according to \eqref{eq:twoqubit-state} with $\bm{n}_{0}=\bm{m}_{0}=[0,0,1/\sqrt{6}]$, $T=\diag\bigl[1/\sqrt{6},-1/\sqrt{6},1/\sqrt{3}\bigr]$. The thin lines mark the final state according to the standard von Neumann projective state update rule.}
    \label{fig:entangled-local}
\end{figure}
\end{widetext}

\vspace{1ex}
\paragraph{Conclusions ---} Employing the notion of quasilinear operations \cite{RembielinskiCaban2020PRR,RembielinskiCaban2021Quantum}, we propose replacing the deterministic part of the von Neumann selective measurement with a corresponding nonlinear yet quasilinear evolution. This replacement is justified because quasilinear evolutions are legitimate quantum-mechanical operations that preserve ensemble equivalence \cite{RembielinskiCaban2020PRR,RembielinskiCaban2021Quantum,AssociatedLong}. As demonstrated in this Letter, the quasilinear evolution equation \eqref{eq:measurement-quasilinear-eq}, which generalizes the von Neumann equation, leads to the same final states as the standard von Neumann projection. This behavior was shown explicitly for two-level quantum systems. Moreover, all essential properties of the projection postulate are retained, including independence of the post-measurement state from the initial state, convergence to a pure state that is an eigenstate of the measured observable, and proper action on entangled states. These properties are insensitive to a wide class of modifications of the $G$ generator, i.e., to the specific measurement procedure. On the other hand, the collapse effect is eliminated, while a continuous and transparent description of the dynamical evolution of the quantum state is introduced. 

From the above, we conclude: If we accept measurement as a quasilinear process over time, then the von Neumann projection postulate simply expresses the final state of the quasilinear evolution (cf.~\cite{JordanSiddiqi2024Book}, Sec.~1.6).

An interesting feature of the proposed framework is the emergence of structural instability, which leads to exceptions to the standard rule in a small region of parameter space very close to configurations with perpendicular vectors $\bm{\omega}$ and $\bm{g}$,  i.e., near the critical angle $\Theta\approx\pm \pi/2$ as well as in trajectory points with $\omega^2=g(t)^2$. This issue and its experimental consequences are discussed in detail in \cite{AssociatedLong}. 

\begin{acknowledgments}
\paragraph{Acknowledgments ---}We thank Pawe{\l} Horodecki for the opportunity to present some of the ideas and results of our work during the main session of the XVI KCIK--ICTQT Symposium ``Quantum Information'', held on 7--10 May 2025 in Gda\'nsk--Sopot, Poland.
\end{acknowledgments}

\bibliography{references}

\end{document}